# Exceptional Points of Degeneracy in Periodic Coupled Waveguides and the Interplay of Gain and Radiation Loss: Theoretical and Experimental Demonstration

Ahmed F. Abdelshafy, Mohamed A. K. Othman, Dmitry Oshmarin, Ahmad T. Almutawa, and Filippo Capolino

*Abstract*— We present a novel paradigm for dispersion engineering in coupled transmission lines (CTLs) based on exceptional points of degeneracy (EPDs). We develop a theory for fourth-order EPDs consisting of four Floquet-Bloch eigenmodes coalescing into one degenerate eigenmode. We present unique wave propagation properties associated to the EPD and develop a figure of merit to assess the practical occurrence of fourth-order EPDs in CTLs with tolerances and losses. We experimentally verify for the first time the existence of a fourth EPD (the degenerate band edge), through dispersion and transmission measurements in microstrip-based CTLs at microwave frequencies. In addition, we report that based on experimental observation and the developed figure of merit, the EPD features are still observable in structures that radiate (leak energy away), even in the presence of fabrication tolerances and dissipative losses. We investigate the "gain and loss balance" regime in CTLs as a mean of recovering an EPD in the presence of radiation and/or dissipative losses, without necessarily resorting to Parity-Time (PT)-symmetry regimes. The versatile EPD concept is promising in applications such as high intensity and power-efficiency oscillators, spatial power combiners, or low-threshold oscillators and opens new frontiers for boosting the performance of large coherent sources.

## I. INTRODUCTION

ELECTROMAGNETIC guiding structures or resonators are characterized by their evolution equations in terms of the eigenmodes (eigenvalues and eigenvectors). Among many features of the evolution of these eigenmodes, we explore the points in the parameter space of such system at which two or more eigenmodes coalesce into a single degenerate eigenmode [1]–[5]. We denote these points as exceptional points of degeneracy (EPD), and the degeneracy order represents the number of coalescing eigenmodes. Periodic guiding structures enable the occurrence of a fundamental class of EPDs at the so-called "regular" band edge (RBE) at which standing waves

This material is based upon work supported by the Air Force Office of Scientific Research award numbers FA9550-15-1-0280 and FA9550-18-1-0355, and by the National Science Foundation under award NSF ECCS-1711975.

The authors are with the Department of Electrical Engineering and Computer Science, University of California, Irvine, CA 92697 USA. (e-mail: abdelsha@uci.edu, mothman@uci.edu, f.capolino@uci.edu).

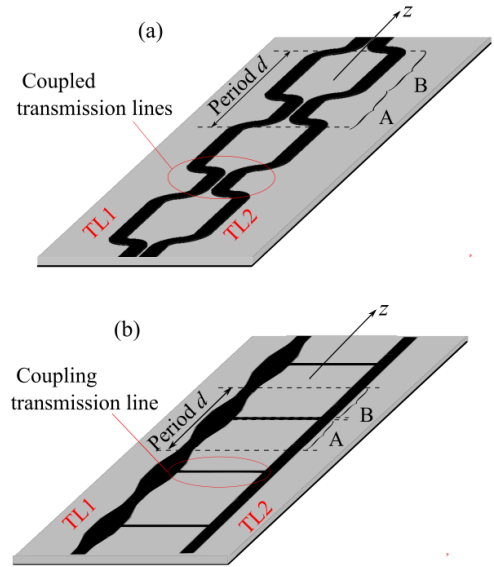

Fig. 1. Example geometries of waveguide microstrip lines on a grounded dielectric slab that support a fourth order EPD, visible in the $(k\text{-}\omega)$ dispersion diagram at microwave frequencies. Examples in (a) and (b) represent two cases of coupling, with proximity fields and with a physical connection, respectively. Results in Sections III and VI are based on a CTL that models the geometry in (a), whereas results in Sections IV and V are based on the microstrip CTL geometry in (b). In the lossless case, the fourth order EPD is called DBE, however such structures exhibit both second and fourth order EPDs in the case of gain and loss balance.

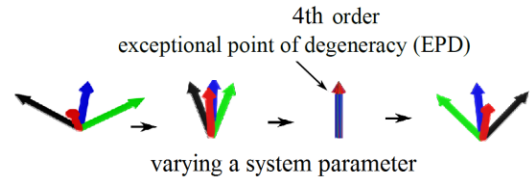

Fig. 2. Representation of the eigenvectors of the coupled waveguides near an EPD, schematically showing that the four-independent vectors coalesce into a degenerate eigenvector at the EPD when one system parameter is varied.

with zero group velocity are manifested at the band edge (i.e., the separation between pass and stop bands). EPDs also occur in Parity-Time (PT)-symmetric structures such as coupled waveguides and resonators when the system's refractive index obeys $n(x) = n^*(-x)$ where $x$ is a coordinate in the system [3], [6]–[8] and * denotes complex conjugation. EPDs in PT-symmetric structures occur in coupled waveguides [1]–[4]. We point out that EPDs cannot occur in systems whose evolution is described by a Hermitian matrix (see Section II). Thus, an EPD occurs in systems where the evolution of the system vector, in space or time, is described by a non-Hermitian matrix which can be imposed by periodicity or by having losses and gain in the system [3], [4]. Moreover, in [9] EPDs have been found in nonreciprocal waveguides, between coupled topological modes, with a balanced distribution of loss and gain. Recently, some of the authors have shown that EPDs of second order can occur with asymmetric distributions of gain and loss in *uniform* CTLs [2]. Furthermore, in [10] second order EPDs in uniform CTLs with loss and gain have been investigated from the bifurcation theory point of view. It is worth noting that an EPD affects the flatness of a $\omega$–$k$ dispersion diagram and in general the higher the EPD order the flatter the dispersion. This concept shall not be confused with the flattening of the group velocity versus frequency dispersion diagram, for instance discussed in [11] and references therein.

In this paper, we investigate second and fourth order EPDs occurring in *periodic* guiding structures whose wave dynamics are represented by two periodic coupled TLs. In particular we are interested in the *fourth order EPDs* that occur when all four independent Floquet-Bloch eigenmodes coalesce in their eigenvalues and eigenvectors and form one single degenerate eigenvector [1], [12], [13] at the band edge as it will be clear later on in the paper, in *absence of losses and gain*. It is worth noting that

A *fourth order EPD* occurring at the band edge of a lossless structure is called degenerate band edge (DBE). This DBE condition is the basis for a possible enhancement of gain in active devices comprising DBE structures [14], [15]. Although the DBE concept was first shown in periodic layers of anisotropic materials where two independent polarizations are coupled throughout the structure [1], [13], other investigations were also carried out using microstrip lines for filtering and antenna applications [16]–[18]. Moreover, coupled silicon waveguides were also designed [19], [20], with potential application in low threshold and high efficiency lasers [14]. However, in connection with experimental studies, the existence of EPD features was shown experimentally in periodic anisotropic layered media consisting of dielectric layers fabricated of low-loss microwave ceramic disks [21]; such geometry exhibits the split band edge [1], [22]. Also, in optical waveguides the giant resonance and giant *Q* factor scaling associated with DBE were observed [20], [23]. Moreover, in printed CTLs frozen guided modes associated to stationary inflection point were observed [24]. More recently, the authors proposed that the DBE can also be manifested in all-metallic periodically-loaded waveguides [25] and experimentally detected even in the presence of losses and tolerances [26]. The application of such waveguide in high power microwave oscillators based on electron beams demonstrates low starting (threshold) current and better threshold scaling with length compared to conventional backward wave oscillators [15], [27], [28]. The same concept has been adopted for investigating ladder circuit oscillators [29], [30] with low threshold and potential power efficiency. Note that the DBE (i.e. fourth order EPD) occurs rigorously only in lossless structures as shown in [1], [14], [16], [25], [26] but fourth order EPDs can also be achieved in gain and loss balanced CTLs as will be shown in this paper. This is different from the CTL regime discussed in [2] since there second order EPDs have been considered in uniform CTLs, while here we deal with *periodic* CTLs and the investigated EPD regimes are of the fourth order. It is also worth noting that the gain compensation condition, the so-called "gain and loss balanced condition" in CTLs investigated in this paper to obtain fourth order EPDs does not necessarily mean PT-symmetric systems. Moreover, in the realization of EPDs with the gain and loss balanced condition, it is important to observe that losses can actually represent distributed radiation in a periodically opened waveguide structure. This concept paves the way for a new class of radiating array oscillators based on EPDs.

Besides the presented general formulation that is applicable to any CTL, we discuss in detail the two microstrip coupled transmission lines shown in Fig. 1. However, the conclusions drawn can be extended to many other geometries or structures since our formalism is general; operating from RF to optical frequencies. The rest of the paper is organized as follows. In Section II we develop the theory of periodic CTLs and discuss the characteristics of EPDs in periodic coupled waveguides. In Section III we introduce the concept of hyperdistance and investigate the effect of losses and coupling on the dispersion diagram as well as the effect of perturbations. In Section IV and V we show the experimental demonstration of EPDs in periodic structures with both infinite- and finite-length structures at microwaves based on coupled microstrip lines. In Section VI we describe EPDs in CTL, based on gain and loss balance.

II. SYSTEM DESCRIPTION OF CTLS

We consider a pair of CTLs such that two independent modal fields are able to propagate along each *z*-direction (so a total of four independent modes are present when we consider propagation in both positive and negative z directions). At an EPD some of these modes (two or four, depending on the case, in a system satisfying reciprocity) will no longer be independent and thus will coalesce. We refer entirely to the formulation in [2] for homogeneous (i.e., uniform) TLs that describe the field evolution using a CTL approach, assuming a time harmonic evolution as $e^{j\omega t}$. In this section we consider the general case of a periodic CTL that may or may not possess dissipative as well as radiation losses and/or gain.

*A. State vector and wave propagation in CTLs*

We start by representing the field amplitudes in the two CTLs by equivalent voltage and current vectors $\mathbf{V}(z) = \begin{bmatrix} V_1(z) & V_2(z) \end{bmatrix}^T$ and $\mathbf{I}(z) = \begin{bmatrix} I_1(z) & I_2(z) \end{bmatrix}^T$. It is convenient to define the four-dimensional state vector

$$\mathbf{\Psi}(z) \equiv \begin{pmatrix} \mathbf{V}(z) \\ \mathbf{I}(z) \end{pmatrix} \quad (1)$$

that comprises voltages and currents at any coordinate *z* in the CTL. This technique has been employed in [1], [15], [31], [32] to investigate the modal properties of photonic crystals and periodic waveguides. The system evolution along *z* is described by the first order differential equations [2]

$$\frac{\partial}{\partial z}\boldsymbol{\Psi}(z) = -j\underline{\mathbf{M}}(z)\boldsymbol{\Psi}(z) \quad (2)$$

where $\underline{\mathbf{M}}$ is a 4×4 CTL system matrix, where $\underline{\mathbf{M}}$ is given by

$$\underline{\mathbf{M}}(z) = \begin{bmatrix} \underline{\mathbf{0}} & -j\underline{\mathbf{Z}}(z) \\ -j\underline{\mathbf{Y}}(z) & \underline{\mathbf{0}} \end{bmatrix} \quad (3)$$

where $\underline{\mathbf{0}}$ is the 2×2 zero matrix, and $\underline{\mathbf{Z}}$ and $\underline{\mathbf{Y}}$ are the series impedance and shunt admittance matrices describing the per unit parameters of the CTL [2], [32], [33]. Note that in [2] we have investigated the uniform CTL case, i.e., the case when $\underline{\mathbf{M}}$ is invariant along z. The properties described in this paper are related to a periodic piece-wise variation of the system matrix $\underline{\mathbf{M}}(z)$.

The series impedance and shunt admittance matrices describing the per unit parameters of the CTL are defined as $\underline{\mathbf{Z}} = j\omega\underline{\mathbf{L}} + \underline{\mathbf{R}}$, and $\underline{\mathbf{Y}} = j\omega\underline{\mathbf{C}} + \underline{\mathbf{G}}$. The CTL per unit length inductance $\underline{\mathbf{L}}$ and capacitance $\underline{\mathbf{C}}$ matrices are 2×2 symmetric and positive definite matrices [31], [34], whereas the per unit length series resistance $\underline{\mathbf{R}}$ and shunt conductance $\underline{\mathbf{G}}$ matrices are 2×2 symmetric matrices [31], [34] accounting for losses and for small-signal linear gain introduced by negative resistance or conductance. Note that $\underline{\mathbf{R}}$ and $\underline{\mathbf{G}}$ are positive-definite matrices if and only if they only represent the losses [31], [34]. Moreover, we recall that the capacitance and conductance matrices have negative off-diagonal entries (see Ch. 4 in [31]). Cutoff conditions could be modeled by resonant series and shunt reactive elements as was done in [15], [32]. For instance, cut-off series capacitances (inductances) per unit length can model cutoff conditions for TM (TE) waves in the waveguide and could be included in the impedance (admittance) matrix (see Ch. 8 in [35]). However, for the sake of simplicity here we ignore cutoff conditions since we analyze microstrip lines in terms of their fundamental quasi-TEM modes that ideally do not have a cutoff frequency.

The homogenous solution of (2), assuming a certain boundary condition at $z = z_0$, namely $\boldsymbol{\Psi}(z_0) = \boldsymbol{\Psi}_0$ inside a *uniform* (i.e., z-invariant) CTL segment, is found by representing the state vector solution at any arbitrary coordinate $z_1$ via

$$\boldsymbol{\Psi}(z_1) = \underline{\mathbf{T}}(z_1, z_0)\boldsymbol{\Psi}(z_0) \quad (4)$$

where we define $\underline{\mathbf{T}}(z_1, z_0)$ as the *transfer matrix* which translates the state vector $\boldsymbol{\Psi}(z)$ between the two points $z_0$ and $z_1$. Within a uniform segment the transfer matrix is easily calculated as

$$\underline{\mathbf{T}}(z_1, z_0) = \exp(-j(z_1 - z_0)\underline{\mathbf{M}}) \quad (5)$$

and the transfer matrix satisfies the group property $\underline{\mathbf{T}}(z_2, z_0) = \underline{\mathbf{T}}(z_2, z_1)\underline{\mathbf{T}}(z_1, z_0)$ and the symmetry property

$$\underline{\mathbf{T}}(z_1, z_0)\underline{\mathbf{T}}(z_0, z_1) = \underline{\mathbf{1}} \quad (6)$$

where $\underline{\mathbf{1}}$ is the 4×4 identity matrix.

The previous discussion identifies the transfer matrix of a uniform segment of a CTL, and it is used to describe periodic waveguides made of a cascade of uniform CTL segments.

### B. Evolution of waves in periodic coupled CTLs

Let us assume a periodic CTL composed of two uniform segments A and B cascaded as shown in Figs. 1(a) and 1(b). The transfer matrix of each individual CTL segment is given by

$$\begin{aligned}\underline{\mathbf{T}}_A &\equiv \underline{\mathbf{T}}(z_0 + d_A, z_0) = e^{-j\underline{\mathbf{M}}_A d_A}, \\ \underline{\mathbf{T}}_B &\equiv \underline{\mathbf{T}}(z_0 + d_B, z_0) = e^{-j\underline{\mathbf{M}}_B d_B}\end{aligned} \quad (7)$$

where, $\underline{\mathbf{M}}_A$ and $\underline{\mathbf{M}}_B$ are defined in terms of the per-unit-length impedance and inductances of the segments A and B, respectively using (3), while $d_A$ and $d_B$ are the lengths of segments A and B respectively. The transfer matrix $\underline{\mathbf{T}}_U$ of the unit cell of the CTL shown in Fig. 1(a), is expressed as the product of the two transfer matrices of the individual segments of the unit cell calculated in (7) as $\underline{\mathbf{T}}_U = \underline{\mathbf{T}}_B\underline{\mathbf{T}}_A$. On the other hand, the transfer matrix of a unit cell of the periodic CTL in Fig. 1(b) incorporates an additional coupling matrix due to the coupling microstrip. Particularly in Fig. 1(b), segments A and B are uncoupled while the coupling between TL1 and TL2 is introduced using a physical connection which is mediated through another transfer matrix denoted by the coupling transfer matrix $\underline{\mathbf{T}}_C$ (whose expression is omitted here for the sake of brevity). Hence, the unit cell transfer matrix is $\underline{\mathbf{T}}_U = \underline{\mathbf{T}}_B\underline{\mathbf{T}}_C\underline{\mathbf{T}}_A$.

The transfer matrix $\underline{\mathbf{T}}_U$ translates the state vector across a unit cell as

$$\boldsymbol{\Psi}(z + d) = \underline{\mathbf{T}}_U\boldsymbol{\Psi}(z) \quad (8)$$

where $d$ is the period in Figs. 1(a) and 1(b). For an infinitely long stack of CTL unit cells, a periodic homogenous solution for the state vector exists in the form $\boldsymbol{\Psi}(z) \propto e^{-jkz}$. This form denotes a Floquet-Bloch type solution where $k$ is the complex Floquet-Bloch wavenumber and $e^{-jkz}$ is referred to as the Floquet-Bloch multiplier. To find such Floquet-Bloch wavenumbers and the eigenvectors, we write the following eigenvalue equation

$$(\underline{\mathbf{T}}_U - \zeta_n\underline{\mathbf{1}})\boldsymbol{\Psi}_n(z) = 0 \quad (9)$$

where $\boldsymbol{\Psi}_n$ are the regular eigenvectors, corresponding to four eigenvalues (or Floquet-Bloch multipliers) given by $\zeta_n = e^{-jk_n d}$, with $n = 1, 2, 3, 4$. The eigenvalues can be readily obtained as solutions of the characteristic equation $\det(\underline{\mathbf{T}}_U - \zeta\underline{\mathbf{1}}) = 0$. We introduce the matrix $\underline{\mathbf{k}}$ as a 2×2 diagonal matrix, whose diagonal elements are the Floquet-Bloch wavenumbers with positive real values, i.e. $\underline{\mathbf{k}} = \text{diag}(k_1, k_2)$. We define the Brillouin zone in our periodic structure from $k = 0$ to $k = 2\pi/d$. Therefore $k_1, k_2, -k_1 + 2\pi/d$, and $-k_2 + 2\pi/d$ are the four modal wavenumbers of the four independent Floquet-Bloch modes in the periodic structure, inside the Brillouin zone. In fact, all the Floquet-harmonics whose wavenumbers are $k_{n,p} = \pm k_n \pm 2p\pi/d$ with $p = 1, 2, 3, \cdots$ obey symmetry because we consider reciprocal systems ($k$ and $-k$ are both solutions). We will also define $\underline{\underline{\Lambda}}$ as a diagonal matrix whose diagonal elements are the eigenvalues $e^{-jk_n d}$ via

$$\underline{\underline{\Lambda}} = \begin{pmatrix} e^{-j\underline{\underline{\mathbf{k}}}d} & \underline{\underline{\mathbf{0}}} \\ \underline{\underline{\mathbf{0}}} & e^{j\underline{\underline{\mathbf{k}}}d} \end{pmatrix} \quad (10)$$

(Note that $e^{-j\underline{\underline{\mathbf{k}}}d}$ is a diagonal matrix with elements $e^{-jk_1 d}$ and $e^{-jk_2 d}$.) Therefore, it follows that the transfer matrix $\underline{\underline{\mathbf{T}}}_U$ is written as

$$\underline{\underline{\mathbf{T}}}_U = \underline{\underline{\mathbf{U}}} \; \underline{\underline{\Lambda}} \; \underline{\underline{\mathbf{U}}}^{-1} \quad (11)$$

where $\underline{\underline{\mathbf{U}}}$ is 4×4 matrix that serves as a non-singular similarity transformation that diagonalizes $\underline{\underline{\mathbf{T}}}_U$, and is computed using the four regular, normalized eigenvectors of $\underline{\underline{\mathbf{T}}}_U$ as $\underline{\underline{\mathbf{U}}} = [\mathbf{\Psi}_1 | \mathbf{\Psi}_2 | \mathbf{\Psi}_3 | \mathbf{\Psi}_4]$ meaning that the column vectors of $\underline{\underline{\mathbf{U}}}$ are the regular eigenvectors of $\underline{\underline{\mathbf{T}}}_U$ that are linearly independent. If (11) is satisfied with a non-singular $\underline{\underline{\mathbf{U}}}$ we say that $\underline{\underline{\mathbf{T}}}_U$ is *similar* to a diagonal matrix.

In principle, the matrix $\underline{\underline{\mathbf{T}}}_U$ is non-Hermitian, but it satisfies some other important properties in the absence of gain and loss, where each constitutive CTL segment has a $z$-evolution matrix $\underline{\underline{\mathbf{M}}}$ that is $J$-Hermitian and a Hermitian characteristic matrix $\underline{\underline{\mathbf{Z}}}\underline{\underline{\mathbf{Y}}}$, mainly the $J$- unitarity property,

$$\underline{\underline{\mathbf{T}}}_U^\dagger(z_1, z_0) = \underline{\underline{\mathbf{J}}} \underline{\underline{\mathbf{T}}}_U^{-1}(z_1, z_0) \underline{\underline{\mathbf{J}}} \quad (12)$$

where $\underline{\underline{\mathbf{J}}}$ is defined in eq. (9) in [2], and the dagger symbol † denotes the complex conjugate transpose operation. Importantly, the constitutive matrices $\underline{\underline{\mathbf{T}}}_A$ and $\underline{\underline{\mathbf{T}}}_B$ for instance, are diagonalizable in the absence of gain and loss in each segment, and therefore both $\underline{\underline{\mathbf{T}}}_A$ and $\underline{\underline{\mathbf{T}}}_B$ have a complete set of eigenvectors. In addition, in the lossless case their eigenvalues lie on the unit circle of the complex eigenvalue plane. The product $\underline{\underline{\mathbf{T}}}_B\underline{\underline{\mathbf{T}}}_A$, on the other hand, is not necessarily diagonalizable and indeed EPDs can manifest in the parameter space of the periodic structure described by $\underline{\underline{\mathbf{T}}}_U$, and the complex Floquet multiplier, which are the eigenvalues of $\underline{\underline{\mathbf{T}}}_U$, may not lie on a circle in the complex eigenvalue plane even in lossless structures as we will show in Section III. We recall that $\underline{\underline{\mathbf{T}}}_U$ is a non-Hermitian matrix which is a necessary condition to realize EPDs, though it is not sufficient by itself since there are non-Hermitian matrices that can be diagonalized. Whereas, the sufficient condition occurs when the matrix $\underline{\underline{\mathbf{T}}}_U$ becomes defective, i.e., similar to a matrix that contains a non-trivial Jordan block, as it will be shown in the next subsection.

*C. Fourth-order exceptional points of degeneracy*

We define the fourth-order EPD as a point in the parameter space of the periodic CTL at which the four eigenvectors coalesce into a single degenerate eigenvector. At an *EPD* the matrix $\underline{\underline{\mathbf{T}}}_U$ becomes defective, i.e., *it cannot be diagonalized*. At a fourth order EPD, the eigenvalue problem (9) does not provide a complete basis of independent eigenvectors: they coalesce into a single one, hence $\underline{\underline{\mathbf{U}}}$ in (11) becomes singular. Strictly speaking, it means that it is not possible to find a non-singular similarity transformation as in (11) that diagonalizes $\underline{\underline{\mathbf{T}}}_U$. It is well-known from linear algebra that an $m \times m$ Jordan block has a single eigenvector. Accordingly, $\underline{\underline{\mathbf{T}}}_U$ must be similar to a matrix that contains a non-trivial Jordan block of order four to have a fourth order EPD. Such a Jordan block will have a single degenerate eigenvector, and three other *generalized* eigenvectors. The four linearly independent generalized eigenvectors in such defective system are found by solving the generalized Floquet-Bloch form

$$[\underline{\underline{\mathbf{T}}}_U - \zeta_d \underline{\underline{\mathbf{1}}}]^q \; \mathbf{\Psi}_q^g(z) = 0, \qquad q = 1, 2, 3, 4 \quad (13)$$

where $\mathbf{\Psi}_1^g(z) \equiv \mathbf{\Psi}_1(z)$ is a regular or ordinary eigenvector while $\mathbf{\Psi}_2^g(z)$, $\mathbf{\Psi}_3^g(z)$ and $\mathbf{\Psi}_4^g(z)$ are the generalized eigenvectors of ranks 2, 3, and 4, respectively (see details of generalized eigenvectors in Ch. 7 in [36]). A fourth order EPD in our CTL system occurs *if and only if* the transfer matrix $\underline{\underline{\mathbf{T}}}_U$ is similar to a Jordan canonical matrix [12], [13] given by

$$\underline{\underline{\mathbf{T}}}_U = \underline{\underline{\mathbf{S}}} \; \underline{\underline{\Lambda}} \; \underline{\underline{\mathbf{S}}}^{-1}, \qquad \underline{\underline{\Lambda}} = \begin{pmatrix} \zeta_d & 1 & 0 & 0 \\ 0 & \zeta_d & 1 & 0 \\ 0 & 0 & \zeta_d & 1 \\ 0 & 0 & 0 & \zeta_d \end{pmatrix} \quad (14)$$

with $\underline{\underline{\mathbf{S}}} = [\mathbf{\Psi}_1^g | \mathbf{\Psi}_2^g | \mathbf{\Psi}_3^g | \mathbf{\Psi}_4^g]$ being composed of one regular eigenvector and three generalized eigenvectors corresponding to a coincident eigenvalue $\zeta_d$ with the multiplicity of four, with $\zeta_d = \exp(-jk_d d)$ and $\underline{\underline{\Lambda}}$ in (14) is a 4×4 Jordan matrix. This is the highest order degeneracy that may be obtained in reciprocal and linear structures with only two coupled TLs, since it combines all supported waves (forward and backward, and/or propagating and evanescent). As the system is reciprocal the $\pm k_d$ symmetry must hold. Therefore if we require to have only one degenerate eigenvalue $\zeta_d = \exp(\pm jk_d d)$ with multiplicity of four, this implies that $k_d$ must be either $k_d = \pi/d$ or $k_d = 0$. Therefore, the fourth order EPD can occur either at the Brillouin zone edge ($k_d = 0$) or at the center ($k_d = \pi/d$). Here we have defined the fundamental Brillouin zone (BZ) within the range from $k_d = 0$ to $2\pi$. The evolution of the four eigenvectors near the fourth order EPD, varying either frequency or any other parameter in the CTL, is schematically depicted in Fig. 2. We also stress that

such an EPD (i.e. 4th order EPD) occurs in an entirely passive structure without gain or loss. Later in Section VI, we show that it may also occur in the presence of gain and loss when a particular *balance* condition is reached. In the following section, we explore examples where three kinds of exceptional points manifest themselves in CTL structures.

### III. FOURTH ORDER EPD IN PERIODIC CTLs

The examples in Fig. 1(a) and (b) illustrate two coupled waveguide geometries that can potentially support a DBE. We recall that the DBE occurs in lossless and gainless structures. Therefore, under this assumption, when the CTL is arranged into a periodic structure like the ones shown in Fig. 1, both $\underline{\mathbf{M}}_A$ and $\underline{\mathbf{M}}_B$ are *J-H*ermitian, and the characteristic matrices $\underline{\underline{\mathbf{ZY}}}$ and $\underline{\underline{\mathbf{YZ}}}$ in (3) are Hermitian. The microstrip geometries in Fig. 1 constitute periodically cascaded segments of coupled/uncoupled microstrip lines that support four Bloch modes (two in each direction). In this section we elaborate on the periodic microstrip TL in Fig. 1(a). The other example in Fig. 1(b) will be utilized in Sections IV and V. In the following, we derive the dispersion relation for the fourth order EPD as well as introduce a figure of merit which we call hyperdistance to assess the quality of such EPD subject to any kind of perturbation, e.g., losses, frequency or structural perturbations like a coupling capacitance. We first consider lossless CTLs while at the end of this section losses in the CTL are also considered when discussing the hyperdistance concept.

#### A. Fourth order degeneracy

We first assume the periodic CTL to be lossless (no gain is introduced yet; we will investigate the case with gain and loss in Section VI), hence the matrix $\underline{\mathbf{M}}$ for each segment satisfies the *J*-Hermiticity property [1] (see also [31, Ch. 6])

$$\underline{\mathbf{M}}^\dagger = \underline{\mathbf{J}}\underline{\mathbf{M}}\underline{\mathbf{J}}^{-1}, \quad \underline{\mathbf{M}} = -\underline{\mathbf{N}}\underline{\mathbf{M}}\underline{\mathbf{N}}^{-1} \quad (15)$$

where $\underline{\mathbf{J}}$ and $\underline{\mathbf{N}}$ are defined in eq. (9) in [2]; and as such each individual CTL segment has four modal solutions that have purely real propagation wavenumbers. Indeed, we recall that each of the constitutive CTL segments has the characteristic matrices $\underline{\underline{\mathbf{ZY}}}$ and $\underline{\underline{\mathbf{YZ}}}$ that are Hermitian and has real eigenvalues as proven in the Appendix of [2].

The periodic CTL has eigenvalues $\zeta_n = \exp(-jk_n d)$, with $n = 1,2,3,4$, where $k_n$ is the Bloch wavenumber, that is obtained from the solution of (9). In order to realize the fourth order EPD at a certain frequency, we impose the Jordan block similarity (14) on the transfer matrix of the periodic unit cell. There are several possible unit cell TL parameters (*L*, *C*) combinations that make $\underline{\mathbf{T}}_U$ satisfies (14). A set that models the geometry in Fig. 1(a) is provided as an example with parameters as in Appendix A, leading to a DBE at 4.03 GHz, which is the case shown in Fig. 3.

#### B. Dispersion perturbation analysis and Puiseux series

When a system parameter of the CTL (a geometry or electrical parameter like $C_m$, or frequency, or because of the introduction of losses) is perturbed by a small parameter $\delta$, the perturbed eigenvalue is written as a perturbation of the degenerate eigenvalue in the neighborhood of an EPD eigenvalue $\zeta_e = \exp(-jk_e d)$ in terms of a fractional power expansion (also called Puiseux series [38]–[40]) in the perturbation parameter $\delta$. Since in this section we focus on the fourth order EPD in the absence of losses and gain (i.e., on the DBE), the eigenvalue is $\zeta_d = \exp(-jk_d d)$ which can be related to CTL wavenumber $k_d$ as

$$k_n(\omega) = k_d + a_1 e^{j(n-1)\frac{2\pi}{4}} \delta^{1/4} + a_2 e^{j(n-1)\frac{2\pi}{2}} \delta^{1/2} + \cdots \quad (16)$$

(the proof is provided in Appendix B). Here $a_n$'s are the fractional series expansion coefficients for the *n*th eigenmodes, and $n = 1,2,3,4$ provide the four possible quartic roots near the EPD. The perturbation factor $\delta$ could be, for example, defined as the normalized detuning of angular frequency from the exact DBE frequency, i.e. $\delta = (\omega_d - \omega)/\omega_d$, where $\omega_d$ is the DBE angular frequency. Note that in (16), we denote $\delta^{1/4}$ as the quartic root in the first quarter of the complex plane since we take the four different quartic roots into account using the exponential terms $\exp(j(n-1)\pi/2)$. Keeping only the lowest order terms in the fractional power expansion in (16) leads to the approximate form of the dispersion near the DBE

$$(\omega - \omega_d) \approx h(k - k_d)^4, \quad (17)$$

where *h* is a parameter that defines flatness of the dispersion near $\omega_d$. Indeed, from (16) and (17), by substituting $\delta = (\omega_d - \omega)/\omega_d$, it is clear that the flatness parameter is $h = \omega_d / a_1^4$, and it depends on the system parameters. The smaller the value of *h* the flatter the dispersion curve, indicating higher *Q* factor [41].

#### C. Dispersion relation for fourth order degeneracy

In Fig. 3 we show the $\omega$-$k$ dispersion diagram for a lossless CTL with circuit parameters in Appendix A, that model the geometry in Fig. 1(a), with three different values of the coupling capacitance per-unit-length $C_m$ (note that the co-diagonal elements of $\underline{\underline{\mathbf{C}}}$ are $-C_m$). In Figs. 3(d, e, f) we only plot the modal curves that are related to modes with purely real wavenumber for the lossless CTL. Figs. 3(a, b, c) show the complex wavenumber evolutions varying frequency. In the following we discuss the occurrence of three possible EPDs in lossless CTL: the RBE, the DBE and split band edge (SBE) by varying $C_m$. We define $C_{m,d}$ as the $C_m$ value at which the DBE occurs. Losses are considered at the end of the section and in the last two columns of Fig. 3 (curves with dashed lines).

In Fig. 3(e) we plot the $\omega$-$k$ dispersion relation of the two modes with purely real wavenumber. Since there are four solutions, for $\omega < \omega_d$ there are two other modes that are evanescent (not shown in Fig. 3(e) but shown in Fig. 3(b)) and all four modes coalesce at $\omega_d$. For $\omega > \omega_d$ all four modes are evanescent. Note that the periodic structure's dispersion diagram is periodic in $\text{Re}(k)$ with period $2\pi/d$, hence we plot the dispersion in the first Brillouin zone defined as $0 \leq \text{Re}[k(\omega)] \leq 2\pi/d$, and because of reciprocity wavenumbers are symmetric about the Brillouin zone center $k = k_d \equiv \pi/d$. The so-called frozen mode regime associated with the DBE is related to the vanishing group velocity and has been discussed in previous publications (see [1], [13]–[15], [24], [25], [27], [32], [42], [43] and references therein). In this section we focus on the characteristic of the fourth order EPD, the DBE, and what happens near it, and in Section VI the effect of gain and loss on the DBE is analyzed.

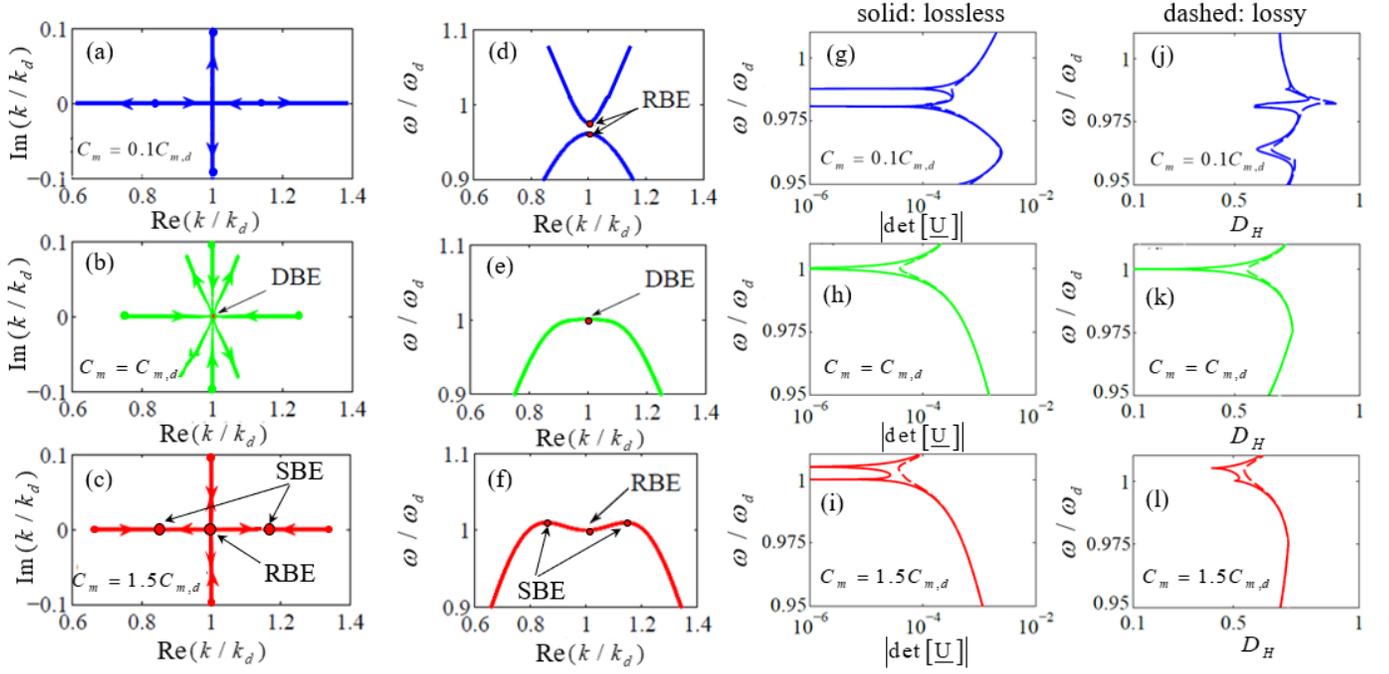

Fig. 3. First two columns: Complex $k$ plane showing the trajectory of the wavenumber $k$ for increasing frequency near the fourth order EPD condition for different values of the coupling capacitances in (a), (b), and (c), for a lossless CTL, and the corresponding dispersion diagram showing the evolution from RBE to DBE and to SBE in (d), (e), and (f), respectively, when varying the coupling capacitance. Dots in the left panel plots indicate the starting point $\omega = 0.95\omega_d$ of the frequency sweep. In (g)-(l) we show the "measures" of the closeness to the EPD calculated based on the four eigenvectors $\Psi_n$ of the lossless and lossy periodic structure: (g), (h), (i) depicts $|\det[\underline{U}]|=0$ while (j), (k), (l) depicts $D_H$ for different values of $C_m$. Any EPD necessarily has $|\det[\underline{U}]|=0$ and a fourth order degeneracy has $D_H(\omega) = 0$. Mathematically speaking, only the lossless case with $C_m = C_{m,d}$ shows a fourth order EPD (i.e., the DBE). Other values of $C_m$ lead to second order EPDs (i.e., the RBEs). Losses perturb the RBE and the DBE. Solid lines represent results for a lossless CTL, whereas dashed lines represent results for a CTL including radiation "loss" in which EPD is no longer strictly manifested, but it can still occur in practice, depending on how small $D_H$ is. Results are based on the microstrip geometry in Fig. 1(a) whose parameters are given in the Appendix A. Here $\omega_e \equiv \omega_d \cong 2\pi(4.03\,\text{GHz})$ and $k_e \equiv k_d = \pi/d$.

The determinant of the similarity transformation $\underline{U}$ of the transfer matrix in (11) is depicted in Figs. 3(g, h, i) varying frequency, for three choices of $C_m$: we observe that $\underline{U}$ is singular ($\det(\underline{U}) = 0$) only at the exceptional points. Note that there exist two different kinds of EPDs: the DBE (fourth order degeneracy, associated to four coalescing eigenvectors) that is the most interesting case and the main subject of the study in this paper, and various RBEs (second order degeneracies associated to two coalescing eigenvectors). In Figs. 3 (a, b, c) we show the evolution of complex wavenumbers in the complex Re($k$)-Im($k$) plane for increasing frequency (the direction of the arrows), for three values of the capacitance per-unit-length $C_m$. By varying $C_m$, three different situations are observed in the dispersion diagrams and they are discussed in the following. The proper coupling capacitance that allows the DBE is denoted as $C_{m,d}$. For small coupling distributed capacitance ($C_m < C_{m,d}$), the CTL exhibits only one lower and one upper RBE at $k = k_d$, as typical in periodic single TL at the band edge. In Fig. 3(a) trajectories of eigenmode coalesce twice at $k = k_d$ at two different frequencies (the upper and lower band edges) designating two distinct second order EPDs. At this second order EPD, two eigenmodes coalesce and this is in principle analogous to the second order EPD that occurs in the uniform CTL investigated in [2] where a balanced gain and loss condition was necessary to develop an EPD. Therefore, a second order EPD can be realized either with a uniform CTL with balanced gain and loss as in [2] or simply by using a periodic TL. In the latter case the EPD occurs at the center of the Brillouin zone, i.e. at $k_d = \pi/d$.

Increasing $C_m$, such that $C_m = C_{m,d}$, we find the proper conditions for the fourth order EPD (the DBE) to be manifested. At the DBE, four Floquet-Bloch eigenmodes coalesce to a single degenerate eigenmode; as explained in Fig. 3(b) where four modal trajectories of complex $k$ varying as a function of frequency intersect at a single point $k = k_d$. At the DBE frequency $\omega_d$ the dispersion diagram in Fig. 3(e) is very flat and thereby approximated by the quartic law (17). Note that this condition cannot occur in *uniform* CTL made of only two TLs with *L* and *C* distributed parameters, as in this paper, since it requires the presence of a bandgap that is endowed by periodicity. (Other important cases with two uniform CTLs with different distributed parameters for backward waves and evanescent waves will be discussed in the future.)

Increasing the coupling such that $C_m > C_{m,d}$ will lead to altering the DBE to a split band edge (SBE) [44]–[46] that has been sometimes referred to as double band edge (DbBE) [18], [42] where three distinct EPDs are found in the dispersion diagram in Fig. 3(f), in the shown frequency range. Each EPD is associated with a second order degeneracy in the eigenmodes as seen in Fig. 3(c). At each EPD the dispersion diagram is flat and it exhibits a vanishing group velocity. It can be observed that when $C_m < C_{m,d}$ or $C_m > C_{m,d}$, two EPDs are manifested, all of the second order, and either all of them on the same side of the stop band (Fig. 3(f)) or on both sides of the stop band (Fig. 3(d)). These EPDs are separated, yet for $C_m = C_{m,d}$ these second order EPDs coalesce and form a fourth order EPD (DBE) (Fig. 3(e)).

### D. Hyperdistance in four-dimensional complex vector space for identifying "vicinity" to a fourth order EPD

In order to distinguish a fourth order EPD among others, and in order to understand how "far" from the EPD a system is, one ought to define a figure of merit (or hyperdistance) to assess the quality of such EPD subject to any kind of perturbation, like losses, frequency detuning, or tolerances of CTL parameters. Here we focus on losses and coupling as the cause of EPD perturbation. Structures, such as that in Fig. 1 are naturally lossy due to dissipative losses (dielectric and ohmic losses) as well as radiation losses that limit the intrinsic quality factors of the constitutive components. Therefore, a perfect degeneracy condition like the DBE corresponding to a lossless structure does not exist in practice when losses are present but can be met in an approximate way and still retain the main features of the four coalescing eigenvectors. In case the periodic CTL is a radiating array, then radiation "losses" (i.e., distributed power extraction) are considered necessary. However, as we will clarify in Section VI, CTLs with (radiation) losses may rigorously exhibit EPDs when a gain is introduced in a *balanced fashion*. Just a few studies have shown the sensitivity of field enhancement in DBE structure to losses [33], [45], [47], and in general any imperfection can be thought as perturbations that affect the eigenvalues (wavenumber) in the way described by (16). Moreover, in designing realistic waveguides, numerical or experimental methods are used and numerical or systematic errors would require a quantitative measure for observing an "exact" EPD. Since our state vector in the CTL is four dimensional, we develop the concept of hyperdistance, denoted as $D_H(\omega)$, *between the four eigenvectors of the transfer matrix* of one-unit cell $\mathbf{T}_U$ to determine the closeness to an EPD. Various choices could be made for its definition, and here the hyperdistance that represents a figure of merit (FOM) is defined as

$$D_H = \frac{1}{6} \sum_{\substack{m=1,n=1 \\ m \neq n}}^{4} \sin(\theta_{mn}), \quad \cos(\theta_{mn}) = \frac{\mathrm{Re}(\boldsymbol{\Psi}_m, \boldsymbol{\Psi}_n)}{\|\boldsymbol{\Psi}_m\| \|\boldsymbol{\Psi}_n\|} \quad (18)$$

with $\theta_{mn}$ representing the angle between two vectors $\boldsymbol{\Psi}_m$ and $\boldsymbol{\Psi}_n$ in a four-dimensional complex vector space with norms $\|\boldsymbol{\Psi}_m\|$ and $\|\boldsymbol{\Psi}_n\|$. Angles are defined via the inner product $(\boldsymbol{\Psi}_m, \boldsymbol{\Psi}_n) = \boldsymbol{\Psi}_m^\dagger \boldsymbol{\Psi}_n$, where the dagger symbol † denotes the complex conjugate transpose operation, and $D_H$ is defined to be always positive. This FOM yields a hyperdistance ideally equal to zero when all four eigenvectors in the CTL system coalesce, i.e., when the CTL system experiences a fourth order EPD. Mathematically this is described by the transfer matrix of the unit cell $\mathbf{T}_U$ becoming similar to a 4×4 Jordan Block as in (14). Therefore, when losses (or any disorder of the structural parameters seen as perturbation) occur, the proposed FOM is not zero. When using numerical methods or measurements (see Sec. IV) we can assume that the EPD is met in practical terms when the FOM measure is less than a very small threshold value, i.e., $D_H < \varepsilon$, where $\varepsilon$ is a small number. It is natural to question when such an EPD occurs in practical terms, i.e., how small $\varepsilon$ shall be to claim an EPD is verified. Furthermore, it is also important to quantify how much losses or perturbations deteriorate the EPD.

To better illustrate this concept, we plot in Figs. 3(j), (k) and (l) the hyperdistance $D_H$ varying as a function of frequency, for a lossless CTL (denoted by solid lines), i.e., for the three values of coupling capacitance $C_m$ considered in Figs. 3(a, b, c). We can now compare the two FOMs introduced: the one associated to vanishing $|\det(\mathbf{U})|$ in Fig. 3(g, h, i) and the one associated to vanishing $D_H$ in Fig. 3(j, k, l). The first one vanishes when at least two eigenvectors coalesce, the second one is the proper one to describe a fourth order EPD because it vanishes only if all four eigenvectors coalesce. Indeed, we see that $|\det(\mathbf{U})| \to 0$ at any EPD, whereas $D_H \to 0$ only at the DBE frequency ($\omega = \omega_d$) and only when $C_m = C_{m,d}$, i.e., only in Fig. 3(k).

The two FOMs are also observed when losses are introduced in the CLT as distributed series resistance (dashed line in Figs. 3(g)-(l)). Losses are assumed to be in both TLs with a quality factor $Q_{\mathrm{TL}}$ of 1000, defined as $Q_{\mathrm{TL}} = \omega L / R$. We also see that the perturbation due to losses can deteriorate the DBE, i.e. $D_H$ is now non-vanishing at $\omega = \omega_d$, when $C_m = C_{m,d}$, in Fig. 3(k). This is in agreement with perturbation theory of eigenmodes leading to (16), implying that a small $\delta$ parameter leads to a significant change in the eigenvalues since $|\delta|^{1/4} \gg |\delta|$ when $|\delta| \ll 1$; which takes place in the close "vicinity" of the fourth order EPD. We again stress that the FOM $D_H$ provides a more quantitative identification of the fourth order degeneracy compared to the simpler measure $|\det(\mathbf{U})|$. The deviation in the ideal EPD conditions due to losses can be seen in Fig. 3(j, k, l) by comparing dashed lines to solid lines for the different kinds of EPDs in Figs. 3(d), (e) and (f) respectively. In a radiating array the limitation in the occurrence of such degeneracies is due *radiation "loss"* (it is natural to introduce gain in the structure to balance the losses and potentially recover the exceptional point as we will show in Section VI). However, the question remains whether we can observe the EPD in experiments even in the presence of losses or not. The concept of hyperdistance developed previously is very useful to decide how "near" a system is to an EPD and therefore if an EPD occurs in practice. Furthermore, the concept of hyperdistance is helpful in determining if introducing gain in some parts of the system is useful in compensating for losses, i.e., in defining the concept of "gain and loss balance" as it will be shown in Section VI.

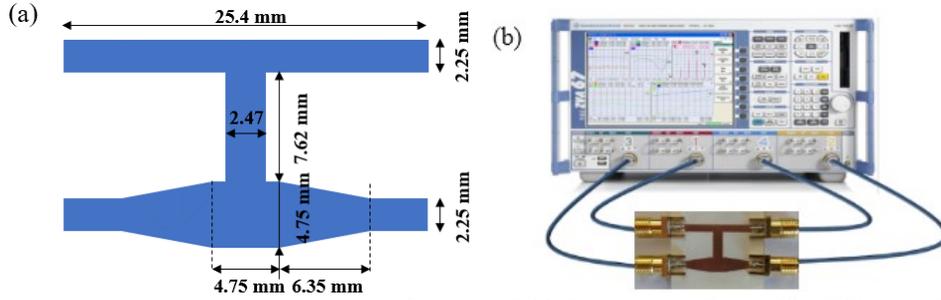

Fig. 4. (a) A microstrip unit cell of a periodic structure that exhibits a fourth order EPD. The substrate is Rogers RO4360G2 with a dielectric constant of 6.15 and height of 1.52 mm. (b) The fabricated unit cell under test with the 4 ports is attached to a VNA to extract the 4-port S parameters versus frequency used to calculate the eigenvalues.

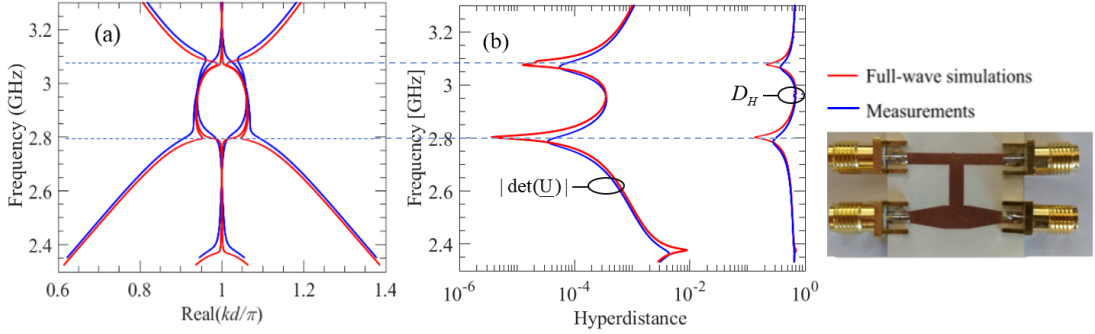

Fig. 5. (a) Dispersion relation and (b) hyperdistance measurement versus full-wave simulation of the unit cell of a periodic CTL exhibiting a $4^{th}$ EPD. The results show that the $4^{th}$ order EPD occurs at 2 different frequencies in the range shown in this plot. Full-wave simulations are performed with Keysight ADS using the method of moments (MoM).

IV. EXPERIMENTAL VERIFICATION OF EPD FEATURES IN PERIODIC CTLs

In this section we experimentally verify the existence of the EPD in the microstrip example in Fig. 1(b). First, instead of fabricating a long (multi-unit cell) coupled microstrip, we experimentally demonstrate the occurrence of EPDs by performing a measurement on a single unit cell fabricated as shown in Fig. 4(a). The unit cell is fabricated on a grounded substrate (Rogers substrate RO4360G2) with a dielectric constant of 6.15, height of 1.52 mm, and with a loss tangent of 0.0038. The TL appearing on the top of the figure was designed to have a characteristic impedance of 50 Ohms and all the unit cell dimensions are reported in Fig. 4(a). To confirm the existence of EPDs in the unit cell, we perform scattering (S)-parameter measurements at the four ports of the unit cell using a four-port Rohde & Schwarz Vector Network Analyzer (VNA) ZVA 67 as shown in Fig. 4(b). From the S-parameters measurements, we then retrieve the transfer matrix $\underline{\mathbf{T}}_U$ as shown from well-known conversion tables [48], [49], for the range of frequencies shown in Fig. 5. The four wavenumbers are derived and plotted in Fig. 5 by solving (9) for complex $k$, at any frequency shown in the plot. The dispersion based on measurements shown in Fig. 5(a) is in a good agreement to the results from full-wave simulations of the S-parameters performed using Keysight Technologies ADS based on the Method of Moments (MoM). The full-wave simulation included the effects of the SMA connectors. The dispersion shows several frequencies at which the four wavenumbers have values very close to each other denoting the occurrence of the $4^{th}$ order EPD, with some perturbation due to ohmic, dielectric and radiation losses. Indeed losses, fabrication tolerances, effect of connectors and other realistic factors affect ideal EPD conditions. Even though mathematically speaking the EPD is not verified exactly, in practical terms, the EPD's prominent features can be well-preserved as we discuss in the following.

In order to assess the existence of the EPD in practical terms, we utilize the hyperdistance concept developed in Section III for the first time. We use the transfer matrix $\underline{\mathbf{T}}_U$ of the unit cell in Fig. 4(a), based on measurements, to calculate the four eigenvectors $\mathbf{\Psi}_n$, with $n = 1,2,3,4$, of the eigenvalue problem in (9). First, the eigenvectors are used to calculate the hyperdistance $D_H$ in (18) that is supposed to vanish when approaching a $4^{th}$ order EPD. Then, they are used to build the similarity transformation matrix $\underline{\mathbf{U}}$ in (10) and hence $|det(\underline{\mathbf{U}})|$ that will tend to zero when at least two eigenvectors become dependent. We plot both $D_H$ and $|det(\underline{\mathbf{U}})|$ FOMs in Fig. 5(b) based on measurements and compare them to the same FOMs obtained from full-wave MoM simulations. We can see that at different frequencies, 2.81 GHz, and 3.1 GHz occurring at $kd = \pi$ the simulated and measured $D_H$ show a minimum denoting vicinity of a $4^{th}$ order EPD despite the presence of radiative and dissipative losses. This demonstrates the EPD properties of eigenvector coalescence is observed at microwaves.

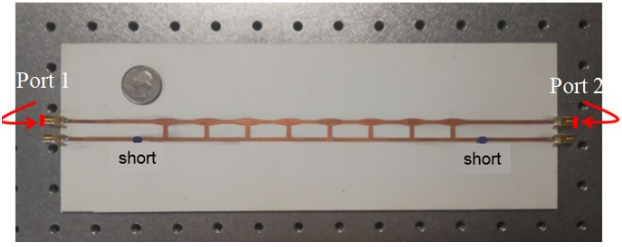

Fig. 6. A fabricated microstrip 8 cell array with an EPD. The CTL is based on the unit cell described in Fig. 4.

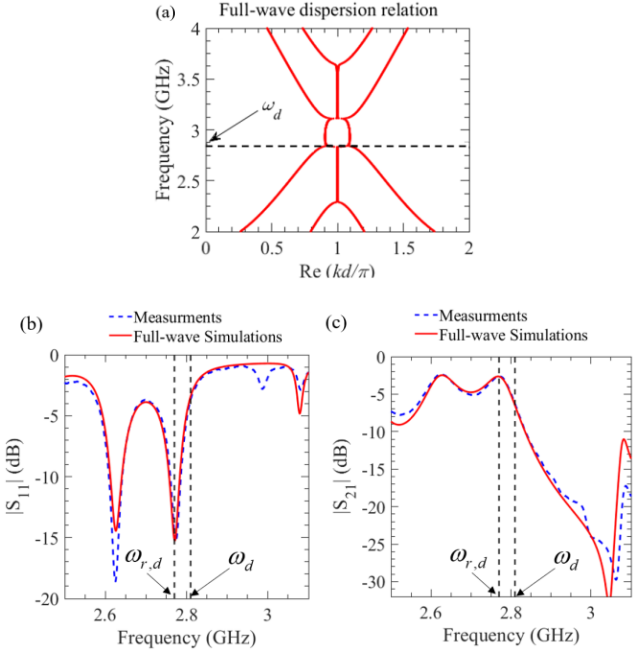

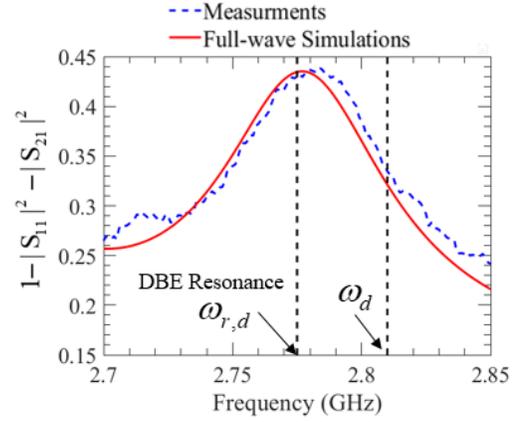

Fig. 8. Measurements versus simulations of the power loss parameter defined as $1-|S_{11}|^2-|S_{21}|^2$ varying as a function of frequency near the EPD resonance.

Fig. 7. (a) Dispersion relation for the unit cell of the periodic structure in Fig. 6 obtained using full-wave finite element method (Ansys HFSS) accounting also for radiation, ohmic and dielectric losses. (b-c) Simulations and measurements of the scattering parameters $S_{11}$ (b) and $S_{21}$ (c) for an 8-unit-cell array in Fig. 6; where a good agreement between full-wave simulations and measurements is shown at the resonance frequency associated with the EPD. The frequency of the resonance peak is estimated by (19).

## V. FINITE LENGTH CTLs WITH EPDs: GIANT RESONANCE AND RADIATION

We investigate the properties of a finite length periodic CTL as in Fig. 6, with a fourth order EPD. In principle, the coalescence of four eigenmodes into one degenerate eigenmode at the DBE causes a quartic dispersion relation (17) and leads to a vanishing group velocity in the infinitely-long periodic structure in the absence of losses. The presence of losses and structural variations/perturbations leads to a non-zero group velocity and renders the hyperdistance (18) non-vanishing as seen from the experiment in Fig. 5. In addition, even in a lossless structure, waves in a *finite-length* CTL with EPD do not have zero-group velocity since the peak of transmission in Fig. 7(c) occurs at a frequency close to the DBE one. Nonetheless, the finite-length CTL develops very interesting resonance features related to the EPDs as we discuss in the following.

Consider the finite length microstrip CTL made of a finite number $N$ of unit cells as shown in Fig. 6. The microstrip and substrate's parameters are the same as those used in relation to Figs. 4 and 5. We perform full-wave simulations using Ansys HFSS utilizing the finite element method (FEM) solver to retrieve the scattering parameters. The dispersion diagram of such structure is shown in Fig. 7(a), obtained by simulating a single unit cell. The 4th order EPD is evident at 2.81 GHz. The 4-port, $N$-cell, finite-length CTL is terminated with coaxial connectors at both extended sides of TL1, the length of the extended part on each side equals 50 mm. Whereas TL2 is terminated in short circuits at the beginning of the 1st unit cell and at the end of the 8th unit cell, as seen in Fig. 6. We are interested in the case where the excited fields inside the CTL create a standing wave because of the Fabry-Pérot cavity (FPC) resonance near the EPD frequency (see details in [1], [14]). At such condition constructive interference of the four synchronous eigenmodes leads to a sharp transmission resonance [1], [14]. Such Fabry-Pérot resonance closest to the DBE frequency occurs at the angular frequency denoted by $\omega_{r,d}$, referred hereafter as DBE resonance, characterized by a peak in the transmission coefficient $|S_{21}|$ and a dip in $|S_{11}|$. The DBE resonance frequency is approximately given by [3],[7]

$$\omega_{r,d} \approx \omega_d - h\left(\frac{\pi}{Nd}\right)^4. \quad (19)$$

Therefore $\omega_{r,d}$ approaches $\omega_d$ for large number $N$ of unit cells. Thanks to the DBE degeneracy condition, a large enhancement in the energy stored is expected because we approach the zero- group velocity conditions when $\omega_{r,d} \approx \omega_d$ and causing a giant enhancement in the $Q$-factor [14], [50].

We show in Fig. 7(b) and (c) FEM full-wave simulations as well as measurement of the magnitude of the scattering parameters $S_{21}$ and $S_{11}$. The full-wave simulation included the effects of the SMA connectors. Results show good agreement between simulation and measurement. Results also demonstrate the occurrence of the DBE resonance at 2.775 GHz, that is close to the DBE frequency of 2.81 GHz. The transmission resonance at the DBE has a transmission coefficient $|S_{21}|$ of $-3.6$ dB.

It is important to point out that such transmission resonance is affected by losses; both dissipative and radiative. Indeed, the exact mathematical DBE condition is not met as seen in Fig. 5 and as discussed in Section IV. Nonetheless, the DBE resonance is still observed in Fig. 7. We emphasize that most of the losses incurred by this CTL are in fact radiation losses. The power loss factor, shown in Fig. 8, is defined as $1-|S_{11}|^2-|S_{21}|^2$ combining both effects of dissipative Ohmic loss as well as radiation. Results show good agreement between FEM full-wave simulations and measurements.

## VI.  EPDs With "Balanced" Gain And Loss

The RBE and DBE (EPDs of order 2 and 4, respectively) exist in structures that are lossless, indeed the presence of losses would inhibit the existence of the DBE, i.e., it would degrade the degree of coalescence of the four eigenmodes. This phenomenon is monitored by the hyperdistance parameter $D_H$ that increases with increasing losses. And this is the reason why we have observed a non-ideal DBE dispersion relation in Fig 5. Increasing radiation losses (e.g., for antenna applications) will further degrade the DBE. Here our aim is to provide a mean to recover EPDs deteriorated by losses by strategically incorporating a proper distribution of gain in each unit cell. The perfect condition that allows the existence of an EPD is named perfect "*loss and gain balance*." (For perfect EPD condition we mean the exact mathematical condition described in Section II that leads to $D_H \to 0$ ). The balancing between gain and loss implies that both the z-evolution matrices $\underline{\mathbf{M}}_A$ and $\underline{\mathbf{M}}_B$ are not *J*-Hermitian, and that the characteristic matrices $\underline{\underline{\mathbf{ZY}}}$ and $\underline{\underline{\mathbf{YZ}}}$ in (3) are non-Hermitian. This does not happen for the DBE case, therefore we will not refer to this condition as DBE. This "loss and gain balance" condition is different from the situation previously studied in PT-symmetric literature [2]–[4], [7], [51] in two aspects: (i) The EPD studied here is of fourth order; (ii) the gain compensation condition for a fourth order EPD in periodic CTLs does not necessarily mean that the system is PT-symmetric [2], [3], [51], [52], i.e., it does not mean that one TL has gain and the other has loss of exact symmetry and magnitude. Note that gain compensation of losses may, in various circumstances, not lead to a perfect EPD as will be shown in the study case B.  However, in many practical cases the *exact* (i.e., perfect) condition may not even be necessary. Indeed, when $D_H$ is sufficiently small we can observe qualitative features related to the EPD even though the structure does not possess the precise mathematical EPD condition. Furthermore, we point out that gain/loss compensation in a unit cell does not necessarily imply absolute instability or self-oscillation in a finite length structure; indeed loading effect must be taken into account in the stability criterion especially near these EPDs as done in [27].

The following discussion is based on the periodic coupled transmission line as in Fig. 1(a), where all parameters are given in Appendix A, investigated with the theoretical tools in Section III. We will study different configuration for losses and incorporating gain.

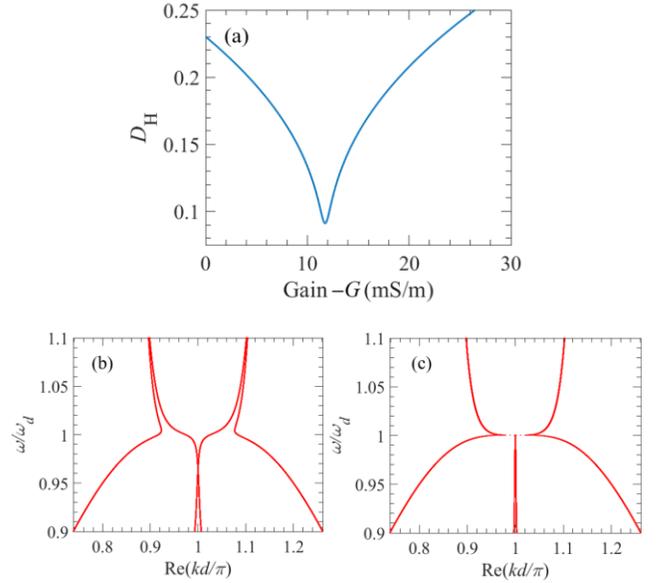

Fig. 9. (a) Plot of the hyperdistance (18) to assess the occurrence of a fourth order EPD in a CTL with losses, as a function of gain $-G$ to compensate for losses. The EPD is achieved in practical terms with $G = G_e = -11.75$ mS/m at which hyperdistance has the minimum value $D_H(\omega_d) = 0.091$ (keeping the angular frequency fixed at $\omega_d$, at which the lossless CTLs develops a DBE). (b) The corresponding dispersion diagram for the lossy CTL (i.e. before introducing gain) where losses are modeled as series resistance $R_n$ such that all TLs have $Q_{TL} = 100$ . (c) The dispersion diagram after introducing the appropriate gain $G = G_e$ to achieve the "balanced gain and loss condition", showing the typical flatness of a DBE is recovered.

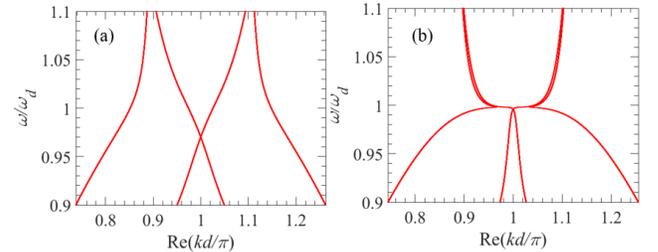

Fig. 10. (a) Dispersion diagram for a lossy CTL structure (i.e. before introducing gain) as in Fig. 9(a) but with one order of magnitude larger losses than in Fig. 9(a), i.e., now all TLs have $Q_{TL} = 10$ . (b) The corresponding dispersion diagram after introducing the gain $G_e = -116.3$ mS/m associated to minimum $D_H(\omega_d) = 0.2$.

*Case A: Series loss and shunt gain*

Losses in the CTL are assumed to be in the form of a per unit length series resistance in both TLs, representing radiation and/or dissipation losses.  Here, gain is introduced in the periodic CTL using a negative per-unit-length conductance $G$ in both TLs (therefore –$G$ is positive with units of Siemens/m). Note that here, we assume that such gain is introduced in both TLs, i.e., each TL has a self-negative conductance $G$ in addition to a series resistance $R_n$ associated to radiation and metal losses for the $n^{th}$ TL, with $n$ = 1,2. Losses are described using the quality factor $Q_{TL,n} = \omega L_n / R_n$ associated to the per-unit-length TL parameters, and here we assume that all TLs have the same quality $Q_{TL,n} = Q_{TL}$ for simplicity. Accordingly, here each constitutive CTL segment has a z-evolution matrix $\underline{\mathbf{M}}$ that doesn't satisfy the *J*-Hermiticity and a characteristic matrix $\underline{\underline{\mathbf{ZY}}}$ that is non-Hermitian. A simple procedure is developed to estimate the value of the needed $G$ in the CTL to "recover" the fourth order EPD occurring at more or less the same angular frequency as that would occur in the lossless structure $\omega = \omega_d$. Note that here for "recovering" an EPD we mean that the required hyperdistance $D_H$ shall be smaller than a specific low threshold $\varepsilon$, that is arbitrarily predetermined (at a lossless 4th order EPD, i.e. at a DBE, this $D_H$ vanishes). Therefore, we find the value of the optimum distributed gain parameter, namely $G_e$ such that the $D_H < \varepsilon$ at $\omega = \omega_d$, where here we assume $\varepsilon = 0.1$ as an example. In Fig. 9 we show the dispersion diagram of a CLT with losses and with compensating gain, as well as the hyperdistance $D_H$ plotted in Fig. 9(a) versus gain parameter $-G$ at $\omega = \omega_d$. We observe that for all TLs with $Q_{TL} = 100$, a chosen values $G = G_e = -11.75$ mS/m, leads to a minimum $D_H < 0.1$ which indicates that the fourth order degeneracy has been achieved in practical terms, i.e., the four eigenvectors $\mathbf{\Psi}_n$ are almost parallel. Indeed, we plot the dispersion relation of the lossy structure (i.e. before introducing gain) in Fig. 9(b). Then we chose the best $G$, i.e. $G = G_e$, and in Fig. 9(c) we plot the dispersion with such gain and loss balance condition. It is clear that the dispersion in Fig. 9(c) is very similar to the ideal case with no loss and no gain plotted in Fig. 3(e).

Additionally, in Fig. 10 we show that this procedure is also useful to recover EPDs when the CTL is strongly perturbed by losses. Indeed in Fig. 10 the CTL and discussion is the same as for the case in Fig. 9, except that now the TLs have losses that are an order of magnitude larger than in Fig. 9, i.e. all TLs considered for the result in Fig. 10 have $Q_{TL} = 10$. It is obvious from Fig. 10(a) that the dispersion of the modes in the lossy CTL (i.e. before introducing gain) is strongly perturbed from the ideal case, indeed the DBE is not visible at all. Then, using the same procedure previously discussed, one can recover the 4th order EPD in practical terms as shown in Fig. 10(b). We recall one more time that these guiding structures are not PT-symmetric since gain and loss are not displaced in a symmetric fashion.

*Case B: shunt loss in both TLs, shunt gain in only one TL*

We analyze another CTL configuration that has losses and gain not satisfying PT-symmetry. Losses in the CTL are assumed to be in the form of a per unit length shunt conductance $G_{L,n}$ in both TLs, representing radiation and/or dissipation losses. Shunt per-unit-length losses are analogously described by a quality factor $Q_{TL,n} = \omega C_n / G_{L,n}$, and in the following we assume that all TLs have shunt conductances $G_{L,n}$ such that $Q_{TL,n} = Q_{TL} = 100$. Gain is introduced in the periodic CTL using a negative per-unit-length conductance $G$ in only TL1 (therefore $-G$ is positive with units of Siemens/m). Note that this CTL configuration differs from the one in Case A in: (i) gain is introduced to only one TL, not in both TLs as in the previous case, and (ii) losses are introduced using a shunt conductance per unit length rather than a series resistance per unit length. Therefore, TL1 has a self-negative conductance $G$ in addition to a $G_{L,1}$ associated to losses, whereas TL2 has only $G_{L,2}$ associated to losses. Similar to Case A, each constitutive CTL segment has a characteristic matrix $\underline{\underline{\mathbf{ZY}}}$ in (3) that is non-Hermitian, and the value of the needed gain $G$ in TL1 to "recover" the fourth order EPD is determined with same procedure used in Case A. In Fig. 11(a) we plot $D_H$ versus the gain parameter $-G$ at $\omega = \omega_d$ (that is the angular frequency at which the DBE for the ideal lossless structure is obtained) and we observe that for $G = G_e = -20.73$ mS/m we have minimum $D_H$ and that such minimum has $D_H < 0.1$ which indicates that the fourth order has been achieved in practical terms. The dispersion relation of the lossy CTL, i.e., before introducing gain, is plotted in Fig. 11(b), and in Fig. 11(c) we show the dispersion after introducing the gain associated to minimum $D_H$. It is worth noting that, the gain compensation leads to a non-ideal EPD. However, in many practical cases the *exact* EPD may not even be necessary. Indeed in Fig. 11(c), one can see that the dispersion diagram is very similar to the one of the ideal DBE, therefore we can still observe qualitative features related to the DBE.

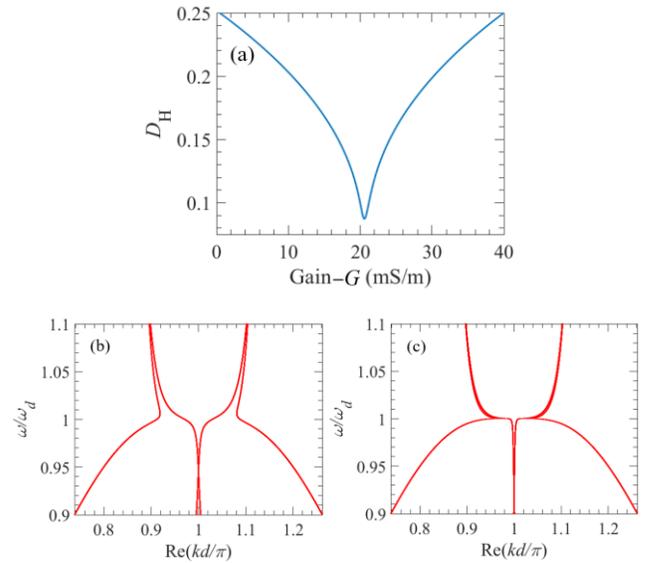

Fig. 11. (a) Plot of the hyperdistance (18) to assess the presence of a fourth order EPD in a CTL with losses, as a function of gain $-G$ to compensate for losses. The EPD is achieved in practical terms with value of gain parameter $-G_e$ = 20.73 mS/m (attached to TL1) at which hyperdistance has the minimum value $D_H(\omega_d) = 0.088$ (i.e., less than the predetermined threshold $\varepsilon$). Angular frequency is fixed at $\omega_d$ at which the lossless CTLs develops a DBE. (b) The corresponding dispersion diagram for lossy structure (i.e. before introducing gain) where losses are modeled as shunt $G_{L,n}$ such that all TLs have same $Q_{TL} = 100$. (c) The corresponding dispersion diagram after introducing the gain associated to minimum $D_H$, showing non-ideal EPD.

Note that, increasing the gain beyond the value of $G = G_e$ may lead to self-oscillations in the unit cell and would render the structure absolutely unstable. Such technique however can be very useful in realizing novel schemes for low threshold oscillators which we will investigate in the near future, and have been already explored in the areas of electron-beam devices [27] and lasers [53].

## VII. CONCLUSION

We have experimentally demonstrated for the first time the occurrence of a fourth order EPD (the DBE) in microstrip CTLs at microwave frequencies, first through four-port measurements of a unit cell leading to the DBE dispersion relation and then through the transmission characteristics of a finite-length CTL. We have also introduced the novel concept of a hyperdistance figure of merit to estimate the effect of perturbations like imperfect coupling, presence of losses and any other perturbation that may arise from fabrications or numerical simulation on the EPD. The hyperdistance concept is a way to measure the eigenvectors mutual "distance" in a multidimensional vector space, that should ideally vanish at the exact EPD condition. The smaller the hyperdistance the closer a system is to the EPD occurrence. Based on the defined figure of merit, our experimental verification has confirmed for the first time that the main EPD features of almost parallel eigenvectors can still exist in realistic periodic arrays that include radiation and dissipative losses. Furthermore, in CTLs that experience significant radiation and dissipative losses, we have shown that the so-called "gain and loss balance" condition leads to recovering an EPD, up to a level that can be quantified via the hyperdistance concept, even in schemes that do not necessarily imply PT-symmetry. The gain and loss balance condition scheme can be applied in principle for any amount of loss in a CTL, paving the way to use it in the design of active grid array antennas.

The capability of obtaining EPDs in CTLs with large radiation losses when adding gain in a proper manner in each unit cell also paves the way to use this multi-eigenmode degenerate scheme in the design of array oscillators and high-intensity spatial power combiners. Potential benefits may include low oscillation threshold, or even high-intensity radiation with high power efficiency, and spectral purity.

## ACKNOWLEDGMENT

The authors would like to thank Prof. A. Figotin, UC Irvine, for very useful discussions. Also, they would like to thank Ansys, Inc. for providing HFSS; Keysight for providing Advanced Design Systems (ADS); and Rogers Corporation for providing the RF laminates.

## APPENDIX A: NUMERICAL PARAMETERS USED IN SECTIONS III & VI

*Periodic CTL parameters used in Section III and VI.* The periodic CTL used has the following parameters pertaining to the microstrip lines shown in Fig. A1(a) over a grounded dielectric slab with a dielectric constant of 2.2 and height of 1.5 mm, which provides a DBE at 4.03 GHz. All microstrips have a width of 1 mm. The coupled line segment has length $d_A = 10$ mm and separation gap 0.2 mm, while the uncoupled segment is made of an uncoupled TL1 with length $d_{B1} = 14$ mm and uncoupled TL2 length $d_{B2} = 10$ mm as shown in Fig. A1(b). Note that the use of the bended junction impacts the assumption of having a uniform segment of a TL1 by slightly increasing the self-inductance and capacitance due to bending [54]. Therefore, the corresponding equivalent CTL parameters of the microstrip line unit cells are as follows. For the coupled section: $C_{11}=C_{22}=$

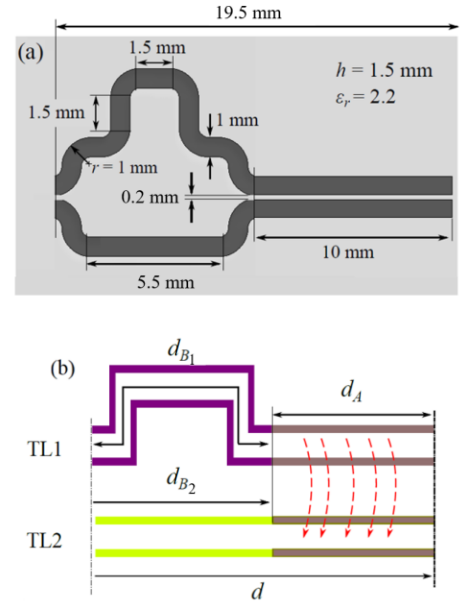

Fig. A1. Geometries of the microstrip configurations adopted in Section III, that develops an EPD. (a), (b) the equivalent passive CTL system of the geometries in Fig. 1(a).

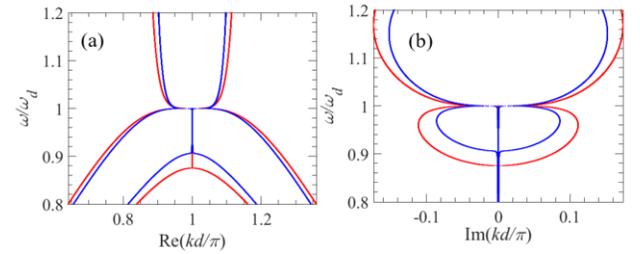

Fig. A2. (a) Real and (b) imaginary parts of the $k$-$\omega$ dispersion diagram obtained for microstrip CTL in Fig. A1(a) using two different methods. First, a circuit simulator (blue curves) adopted in Keysight Technologies ADS. Secondly, by solving eq. (9) for the equivalent passive CTL system shown in Fig. A1(b) (red curves), whose circuit parameters ($L$, $C$) are stated in this Appendix. In both simulations, losses have been neglected.

55.15 pF/m, $L_{11}=L_{22}=0.467$ μH/m, $C_{12} = C_{21}= -C_m = -27.2$ pF/m, $L_{12}= L_{21}=L_m=0.25$ μH/m, are the matrix entries of the $\underline{\underline{C}}$ and $\underline{\underline{L}}$ matrices. For the uncoupled section: TL1 has $L = 0.54$ μH/m, and $C = 42.86$ pF/m whereas TL2 has $L = 0.5$ μH/m, and $C = 35$ pF/m. Most of the results in Section III are for lossless CTLs, losses are considered at the end of Sec. III and in the examples in Section VI and in the full-wave simulation results shown in Fig. A3 for the structure in Fig. A1(a).

In Fig. A2, we show the dispersion diagram (for both real and imaginary parts of $k$-$\omega$) using circuit simulator, which is based on predefined models for each piece of the circuit and using those models the system response is described by a system of equations solved using implicit integration methods, adopted in Keysight Technologies ADS (denoted by blue curves). Also, we plot the dispersion obtained using eq. (9) for the passive equivalent circuit whose circuit parameters mentioned before (denoted by red curves). Both show a good agreement around the DBE point which is clearly observed at $f_d$ =4.03 GHz.

Additionally, in Fig. A3 we plot the real and imaginary parts of the dispersion using MoM full-wave solver adopted in

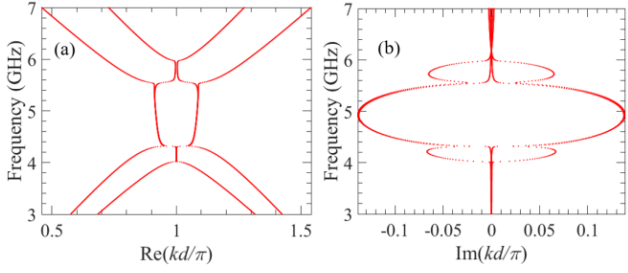

Fig. A3. (a) Real and (b) imaginary parts of dispersion diagram obtained for microstrip CTL in Fig. A1(a) using MoM full-wave simulation (Keysight Technologies ADS) accounting for radiation, ohmic and dielectric losses.

Keysight Technologies ADS. Losses considered in this full-wave simulation includes radiation, ohmic (i.e. copper has conductivity $\sigma = 5.8 * 10^7$ Siemens/m), and dielectric (i.e. substrate used with dielectric constant of 2.2, height of 1.5 mm, and with a loss tangent of 0.002) losses. The dispersion shows a deformed flat dispersion at frequency 4.3 GHz which is slightly upshifted from the DBE frequency $f_d$ obtained for the lossless case using circuit simulator and equivalent passive model in Fig. A2.

### APPENDIX B: WAVENUMBER PERTURBATION IN THE VICINITY OF THE DBE

We show here the relation between the expansion of perturbed eigenvalues of periodic CTL $\zeta_n = \exp(-jk_n d)$ and the perturbed wavenumbers $k_n$, when a system parameter of the CTL is perturbed by a small perturbation parameter $\delta$. In general, the perturbed eigenvalues $\zeta_n$ are written as a perturbation of the degenerate eigenvalue in the neighborhood of the DBE eigenvalue $\zeta_d = \exp(-jk_d d)$ in terms of a fractional power expansion (also called Puiseux series [38]–[40]) in the perturbation parameter $\delta$ as

$$\zeta_n(\delta) = \zeta_d + \alpha_1 e^{j(n-1)\frac{2\pi}{4}} \delta^{1/4} + \alpha_2 e^{j(n-1)\frac{2\pi}{2}} \delta^{2/4} + \cdots \quad (B1)$$

where $\alpha_n$'s are the coefficients of the fractional series expansion of the $n$th eigenmodes, and $n = 1,2,3,4$ provide the four possible quartic roots near the DBE. The perturbed wavenumber $k_n = jd^{-1}\ln(\zeta_n)$ can be obtained as well, by expanding $\ln(\zeta_n)$ around $\zeta_n = \zeta_d$ using the Taylor series expansion which is given by

$$\ln(\zeta_n) = \ln(\zeta_d) + \frac{\zeta_n - \zeta_d}{\zeta_d} - \frac{(\zeta_n - \zeta_d)^2}{2\zeta_d^2} + \cdots. \quad (B2)$$

Accordingly, the perturbed wavenumber $k_n$ is written as a perturbation of the degenerate wavenumber in the vicinity of the DBE eigenvalue $k = k_d$ as follows

$$k_n(\delta) = k_d + a_1 e^{j(n-1)\frac{2\pi}{4}} \delta^{1/4} + a_2 e^{j(n-1)\frac{2\pi}{2}} \delta^{1/2} + \cdots \quad (B3)$$

where

$$a_1 = jd^{-1} e^{jk_d d} \alpha_1$$

$$a_2 = jd^{-1} e^{jk_d d} \left( \alpha_2 - \frac{\alpha_1^2 e^{jk_d d}}{2} \right). \quad (B4)$$

### APPENDIX C: PROPERTIES OF SYSTEM AND TRANSFER MATRICES OF PERIODIC COUPLED TRANSMISSION LINES

We summarize here some important properties associated with matrices related to the system description for the coupled transmission lines.

*First*, a system description is done by defining a state vector $\underline{\Psi}(z)$ as in (1) that varies in a uniform segment as shown in (2) where $\underline{\underline{M}}$ is the constant system matrix. This system matrix entries in (3) are the impedance $\underline{\underline{Z}} = j\omega\underline{\underline{L}} + \underline{\underline{R}}$ and admittance matrices $\underline{\underline{Y}} = j\omega\underline{\underline{C}} + \underline{\underline{G}}$. It is clear from this representation that $\underline{\underline{Z}}$ and $\underline{\underline{Y}}$ are skew-Hermitian matrices, i.e., $\underline{\underline{A}}^\dagger = -\underline{\underline{A}}$, if and only if they are lossless/gainless (i.e. $\underline{\underline{Z}} = j\omega\underline{\underline{L}}$ and $\underline{\underline{Y}} = j\omega\underline{\underline{C}}$, where $\underline{\underline{L}}$ and $\underline{\underline{C}}$ are 2×2 symmetric and positive definite matrices [31], [34]). However, introducing gain and/or loss to the system makes $\underline{\underline{Z}}$ and $\underline{\underline{Y}}$ non-Hermitian. *Therefore*, if we consider only lossless/gainless systems, both $\underline{\underline{ZY}}$ and $\underline{\underline{YZ}}$ matrices are Hermitian (i.e. $\underline{\underline{A}}^\dagger = \underline{\underline{A}}$) as they are real symmetric matrices, which implies that they are diagonalizable and they have real eigenvalues and real determinant [36].

*Second*, the system matrix $\underline{\underline{M}}$ in a lossless/gainless system described by (2), which has the skew-Hermitian matrices $\underline{\underline{Z}}$ and $\underline{\underline{Y}}$ as its entries (3), is a non-Hermitian matrix as

$$\underline{\underline{M}}^\dagger = \underline{\underline{M}}^T \neq \underline{\underline{M}} \quad (C1)$$

but it satisfies the J-Hermiticity property (i.e., $\underline{\underline{A}}^\dagger = \underline{\underline{J}}\underline{\underline{A}}\underline{\underline{J}}$, where $\underline{\underline{J}}$ is any matrix satisfying $\underline{\underline{J}} = \underline{\underline{J}}^{-1} = \underline{\underline{J}}^\dagger$ [34]). This property is easily proved by applying the Hermitian operator to the $\underline{\underline{JM}}$ matrix, i.e., by evaluating the complex conjugate transpose $(\underline{\underline{JM}}(z))^\dagger = \underline{\underline{M}}^\dagger(z)\underline{\underline{J}}^\dagger$; then using (C1) it follows that

$$(\underline{\underline{JM}}(z))^\dagger = \underline{\underline{M}}^\dagger(z)\underline{\underline{J}}^\dagger = \underline{\underline{J}}\underline{\underline{J}}\underline{\underline{M}}^T(z)\underline{\underline{J}} = (\underline{\underline{JM}}). \quad (C2)$$

Accordingly, $\underline{\underline{JM}}$ is a Hermitian matrix, hence $\underline{\underline{M}}$ satisfies the J-Hermiticity property $\underline{\underline{M}}^\dagger = \underline{\underline{JMJ}}$.

*Third*, the periodic structure consists of at least two uniform segments A and B, and the state vector $z$-variation is described using the transfer matrix concept for each unit cell $\underline{\underline{T}}_U$ as in (8), that is the product of the transfer matrices of the two segments $\underline{\mathbf{T}}_U = \underline{\mathbf{T}}_B \underline{\mathbf{T}}_A$, each one given in (7). Within a uniform segment (either A or B), the system vector varies as in (4) and the transfer matrix is given by $\underline{\mathbf{T}}(z) = \exp(-j\underline{\mathbf{M}}z)$ as shown in in (5), where the constant $\underline{\mathbf{M}}$ matrix at the exponent satisfies the J-Hermiticity properties as shown in (C2). Using the following expansion that defines the transfer matrix

$$\underline{\mathbf{T}}(z) = \exp(-j\underline{\mathbf{M}}z) = \mathbf{1} + (-j\underline{\mathbf{M}}z) + \frac{(-j\underline{\mathbf{M}}z)^2}{2!} + \cdots \quad (C3)$$

the spatial variation is written as

$$\begin{aligned}\partial_z \underline{\mathbf{T}}(z) &= \partial_z (\mathbf{1} + (-j\underline{\mathbf{M}}z) + \frac{(-j\underline{\mathbf{M}}z)^2}{2!} + \cdots \\ &= -j\underline{\mathbf{M}} + \frac{(-j\underline{\mathbf{M}})^2 2z}{2!} + \frac{(-j\underline{\mathbf{M}})^3 3z^2}{3!} + \cdots \\ &= -j\underline{\mathbf{M}}(\mathbf{1} + (-j\underline{\mathbf{M}}z) + \frac{(-j\underline{\mathbf{M}}z)^2}{2!} + \cdots \\ &= -j\underline{\mathbf{M}}\underline{\mathbf{T}}(z).\end{aligned} \quad (C4)$$

Accordingly, the transfer matrix is proven to satisfy the J-unitarity (i.e., $\underline{\mathbf{A}}^\dagger = \underline{\mathbf{J}}\underline{\mathbf{A}}^{-1}\underline{\mathbf{J}}$) following the proof in Appendix A of [12], based on the fact that $\partial_z[\underline{\mathbf{T}}(z)\underline{\mathbf{T}}^{-1}(z)] = (\partial_z \underline{\mathbf{T}})\underline{\mathbf{T}}^{-1} + \underline{\mathbf{T}}(\partial_z \underline{\mathbf{T}}^{-1}) = 0$ and the properties shown in (C2) and (C4).

*Finally*, we prove that the unit-cell transfer matrix $\underline{\mathbf{T}}_U = \underline{\mathbf{T}}_A \underline{\mathbf{T}}_B$, which translates the state vector across a periodic unit-cell is J-unitarity. Indeed, as just proved, both $\underline{\mathbf{T}}_A$ and $\underline{\mathbf{T}}_B$ are J-unitarity. Hence, $\underline{\mathbf{T}}_U$ also satisfies the J-unitarity property as it can be easily proven as follows:

$$\begin{aligned}\underline{\mathbf{T}}_U^\dagger &= (\underline{\mathbf{T}}_A \underline{\mathbf{T}}_B)^\dagger = \underline{\mathbf{T}}_B^\dagger \underline{\mathbf{T}}_A^\dagger \\ &= \underline{\mathbf{J}}\underline{\mathbf{T}}_B^{-1}\underline{\mathbf{T}}_A^{-1}\underline{\mathbf{J}} = \underline{\mathbf{J}}\underline{\mathbf{T}}_U^{-1}\underline{\mathbf{J}}.\end{aligned} \quad (C5)$$